\documentclass[10pt]{article}

\usepackage{epsfig}
\usepackage{graphicx}
\usepackage{color}
\usepackage{xspace}
\usepackage{fancyhdr}
\def\lspace{1.0}

\def\npdg{$\overrightarrow{n} + p \rightarrow d + \gamma$}

\def\npdg{$\vec{n} + p \rightarrow d + \gamma$}
\def\BC{${\rm B_4C}$}
\def\CCL{${\rm CCl_4}$}

\def\Gr{$\gamma$-ray}
\def\Grs{$\gamma$-rays}

\def\npdg{$\overrightarrow{n} + p \rightarrow d + \gamma$}

\title{ {\large ICANS-XVII \\
17th Meeting of the International Collaboration on Advanced Neutron Sources \\
April 25-29, 2005 \\
Santa Fe, New Mexico }\\
{\large {\bf The NPDGamma Experiment at LANSCE}}}

\author{{\normalsize M.T.~Gericke$^{c,~e}$, 
J.D.~Bowman$^{a}$, 
R.D.~Carlini$^{j}$,
T.E.~Chupp$^{h}$, 
K.P.~Coulter$^{h}$} \\ 
{\normalsize M.~Dabaghyan$^{d}$,
M.~Dawkins$^{b}$, 
D.~Desai$^{g}$, 
S.J.~Freedman$^{l}$,
T.R.~Gentile$^{m}$,
R.C.~Gillis$^{c}$, }\\
{\normalsize G.L.~Greene$^{f,~g}$,
F.W.~Hersman$^{d}$,
T.~Ino$^{k}$,
G.L.~Jones$^{n}$,
M.~Kandes$^{h}$,
B.Lauss$^{l}$, } \\
{\normalsize M.B.~Leuschner$^{b}$,  
W.R.~Lozowski$^{b}$,
R.~Mahurin$^{g}$,
Y.~Masuda$^{k}$, 
M.~Mason$^{d}$
G.S.~Mitchell$^{a}$ } \\
{\normalsize S.~Muto$^{k}$,
H.~Nann$^{b}$, 
S.A.~Page$^{c}$,
S.I.~Penttil{\"a}$^{a}$,  
W.D.~Ramsay$^{e}$,
S.~Santra$^{b}$, 
P.-N.~Seo$^{p}$, }\\
{\normalsize E.I.~Sharapov$^{o}$,
T.B.~Smith$^{i}$,
W.M.~Snow$^{b}$, 
W.S.~Wilburn$^{a}$, 
V.~Yuan$^{a}$,
H.~Zhu$^{d}$}}
\date{}

\begin{document}
%\linespread{1}
\maketitle
{\normalsize
\begin{center}
  $^{a}$Los Alamos National Laboratory, Los Alamos, New Mexico 87545, USA 

  $^{b}$Indiana University, Bloomington, Indiana 47405, USA

  $^{c}$University of Manitoba, Winnipeg, Manitoba R3T 2N2, Canada

  $^{d}$University of New Hampshire, Durham, NH 03824, USA

  $^{e}$TRIUMF, 4004 Wesbrook Mall, Vancouver, British Columbia V6T 2A3, Canada

  $^{f}$Oak Ridge National Laboratory, Oak Ridge, TN 37831, USA

  $^{g}$University of Tennessee, Knoxville, TN 37996, USA

  $^{h}$University of Michigan, Ann Arbor, MI 48104, USA

  $^{i}$University of Dayton, Dayton, OH 45469, USA

  $^{j}$Thomas Jefferson National Accelerator Facility, Newport News VA 23606,USA

  $^{k}$High Energy Accelerator Research Organization (KEK), Tukubash-shi, 305-0801, Japan

  $^{l}$University of California at Berkeley, Berkeley CA 94720-7300, USA

  $^{m}$National Institute of Standards and Technology, Gaithersburg, MD 20899-0001, USA

  $^{n}$Hamilton College, Clinton, NY 13323, USA

  $^{o}$Joint Institute for Nuclear Research, Dubna, Russia

  $^{p}$North Carolina State University, Raleigh, NC 27695, USA
\end{center}}
%\linespread{\lspace}
\begin{abstract}
\noindent
{\em The NPDGamma collaboration has constructed and commissioned an
apparatus to determine the size of the pion-nucleon coupling constant
in the parity non conserving pion exchange weak potential for N-N
interactions. This coupling constant is directly proportional to the
parity violating up-down asymmetry in the angular distribution of
\Grs~with respect to the neutron spin direction in the capture
of polarized cold neutrons on protons. The measurement of the weak
${\rm \pi NN}$ coupling will provide a test for the effective theory,
describing the nucleon-nucleon interaction as mediated by the exchange
of mesons, and provide results against which to compare models
describing QCD at low energy. NPDGamma is located at the Los Alamos
Neutron Science Center (LANSCE) and utilizes the special features of
cold spallation neutrons to make it possible to measure very small
\Gr~ asymmetries.  In this paper, we present the motivation for the
experiment and report on experimental setup as well as the current
status of the project and the results obtained during the 2004
commissioning run including parity violation asymmetry measurements on
Al, \CCL, In, \BC, and Cu.}
\end{abstract}

\section{Introduction}
\label{scn:INSU}
Since 1980, the weak parity-violating nucleon-nucleon interaction has
typically been described by a meson-exchange potential involving seven
weak meson-nucleon coupling constants~\cite{pp:ddh}. The weak
interaction changes the parity and isospin ($\Delta I = 0, 1, 2$) of
the nucleon-nucleon pair and perturbatively introduces parity
violating admixtures in nuclear wave functions. The study of the
hadronic weak interaction is of great relevance for low energy,
non-perturbative QCD. The hadronic weak couplings probe short range
correlations between quarks because the quark-quark weak interaction
occurs when the distance between quarks is $\leq 2\times10^{-3}~{\rm
fm}$. The electro-weak Standard Model predicts that charged current
contributions to the weak N-N interaction are suppressed and that,
therefore, the measurement of a zero asymmetry in the \npdg~ reaction
would suggest that neither neutral currents nor the strange quark pair
sea contribute significantly to the hadronic weak interaction. A
non-zero \npdg~ asymmetry, on the other hand, would then establish
weak neutral currents as the dominant factor with possibly significant
contributions from strange quarks.

The NPDGamma experiment is under commissioning at the Los Alamos
Neutron Scattering Center (LANSCE). It is the first experiment
designed for the new pulsed high flux cold beam line, flight path 12,
at LANSCE.  NPDGamma will determine the very small weak pion-nucleon
coupling constant $f_{\pi}$ in the nucleon-nucleon
interaction~\cite{pp:prop,pp:snow,pp:snow2,pp:greg}.  This coupling
constant is directly proportional to the parity-violating up-down
asymmetry $A_{\gamma}$ in the angular distribution of 2.2-MeV \Grs~
with respect to the neutron spin direction (eqn.~\ref{eqn:CRSS})
in the reaction \npdg.
\begin{equation}
  \frac{d\sigma}{d\Omega}\propto \frac{1}{4\pi}
  \left(1+A_{\gamma}\cos\theta\right) \label{eqn:CRSS}
\end{equation}
The asymmetry has a predicted size of $5 \times 10^{-8}$ and our goal
is to measure it to 10\% accuracy.  The small size of the asymmetry and the
high proposed measurement precision impose heavy requirements on the
performance of the beam line and apparatus.  It is necessary to
achieve high counting statistics while at the same time suppressing
any systematic errors below the statistical limit. The experiment was
designed to satisfy these requirements~\cite{pp:prop}.

During commissioning the radiative neutron capture on various target
materials was investigated to look for any asymmetry which may enter
as a systematic effect while taking data with the hydrogen target. The
measurements concentrated on materials that can be found within the
experimental apparatus and which are interacting with the
neutrons. The targets included Al, Cu, In, and \BC.  In addition, the
known asymmetry in \CCL~was used to demonstrate that the array
functions as designed and is capable of measuring non-zero
asymmetries. Boron is used throughout the experiment, for neutron
shielding and to collimate the beam. Aluminum is used in most of the
equipment and the beam encounters several millimeter of it, primarily
in the windows of the hydrogen target. Cu and In are also used in the
target. It is therefore necessary to establish the size of the
$\gamma$-asymmetry due to neutron capture on each of these elements.

\section{The Experiment}

The NPDGamma experiment is located on flight path 12 at the Manuel
Lujan Jr. Neutron Scattering Center at LANSCE. The statistical
accuracy that can be reached in the NPDGamma experiment at LANSCE is
limited by the available cold neutron flux.  In a spallation neutron
source, the neutron flux depends on the proton current, the energy
incident on the spallation target, the moderator performance
(brightness) and the neutron guide performance~\cite{pp:pil}.  The
LANSCE linear accelerator delivers $800$~MeV protons to a proton
storage ring, which compresses the beam to $250$~ns wide pulses at the
base, at a rate of $20$~Hz. The protons from the storage ring are
incident on a split Tungsten target and the resulting spallation
neutrons are cooled by and backscattered from a cold ${\rm H_2}$
moderator with a surface area of $12 \times 12$~${\rm cm^2}$ .

\subsection{Beam Line}
\begin{figure}[h]
  \hspace{1cm}
  \includegraphics[scale=0.35]{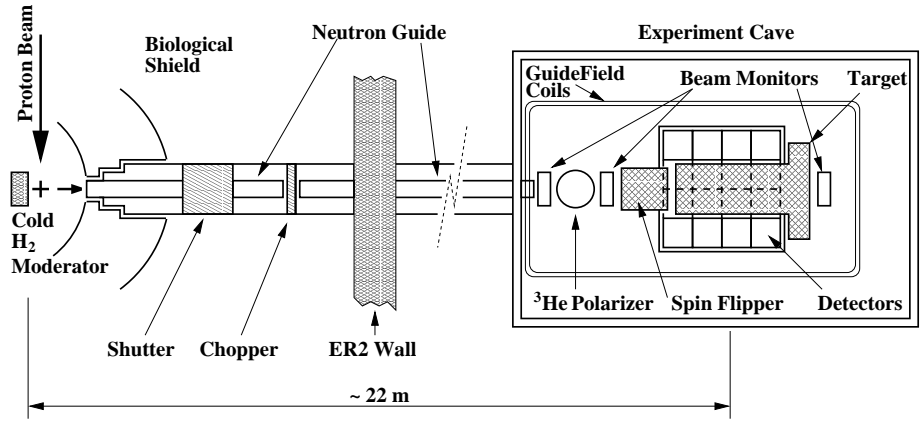}
  \caption{{\em Schematic of the NPDGamma experimental setup.}} \label{fig:EXPSTP}
\end{figure}

Figure~(\ref{fig:EXPSTP}) illustrates the flight path and experiment
cave.  The distance between the moderator and target is about $22$
meters.  The flight path 12 beam line is $\approx 19.5$~m long and
consists of $4$~m of in pile guide, a $2$~m long shutter, a frame
overlap beam chopper and $\approx 13$ meters of neutron guide. The
pulsed neutron source allows us to know the neutron time of flight or
energy and the installed beam chopper allows us to select a range of
neutron energies.  In the experiment cave, the beam is transversely
polarized by transmission through a polarized $^{3}$He cell. Three
$^{3}$He ion chambers are used to monitor beam intensity and measure
beam
polarization through transmission ratios. A radio frequency spin
flipper is used to reverse the neutron spin direction on a pulse by
pulse basis. For the production experiment the neutrons will capture
in a 21 liter liquid para-hydrogen target.  The \Grs~from the
capture are detected by an array of 48 CsI(Tl) detectors operated in
current mode~\cite{pp:darr,pp:swb}. The entire apparatus is located in
a homogeneous 10 Gauss with a field gradient of less than $1$~mG/cm
which is required to maintain the neutron spin downstream of the
polarizer.

For the ideal experiment, with 100\% beam polarization, 100\% spin
flip efficiency, point sources and detectors, and no systematic
effects resulting in background asymmetries or beam depolarization,
the yield from a single detector is given by Eqn.~\ref{eqn:CRSS}, with
the intensity essentially determined by the neutron flux.  However,
the actual signal in the detectors is more closely related to the
number of neutrons that capture on the target, the target size, the
neutron capture location in the target along the beam direction and
the average detector solid angle.  All components between the guide
exit and the target have been optimized to ensure that they can
properly function while attenuating the beam as little as
possible. This includes optimization of the monitor and spin filter
thicknesses, reduction of the aluminum windows on the monitors and the
spin flipper and the overall reduction of the experiment length to
reduce beam divergence.  In addition to maximizing the number of
captured neutrons, a successful asymmetry measurement requires a
stable, polarized beam and the ability to reverse the beam
polarization without significant losses.  The emitted \Grs~ have to be
detected with high efficiency and with reasonably good angular
resolution, which is limited by the finite size of the detectors and
targets~\cite{th:ger}.

The FP12 beam guide was installed to deliver the maximum possible
number of low energy neutrons to the experimental apparatus.  FP12
uses an m=3 guide with a $9.5 \times 9.5$~${\rm cm^2}$ cross-sectional
area.  The guide is coated with hundreds of layers of ${\rm ^{58}Ni}$
and ${\rm ^{47}Ti}$.  It allows neutrons with 3 times the normal
perpendicular velocity to be transmitted, resulting in a large
increase in neutron flux as compared to standard guides which employ
only ${\rm ^{58}Ni}$ coating.  A detailed description of our
measurement of the FP12 moderator brightness and performance of the
neutron guide is given in~\cite{pp:pil}. The measured brightness has a
maximum of $1.25 \times 10^8$ ${\rm n/(s-cm^{2}-sr-meV-\mu A)}$ for
neutrons with an energy of $3.3$~meV.
%$1.25 \times 10^8$ ${\rm n/s/cm^{2}}/{\rm sr/meV/\mu A}$

\subsection{Beam Chopper}
\begin{figure}[h]
  \hspace{3.5cm}%bb=100 0 600 800,clip,
  \includegraphics[scale=0.3]{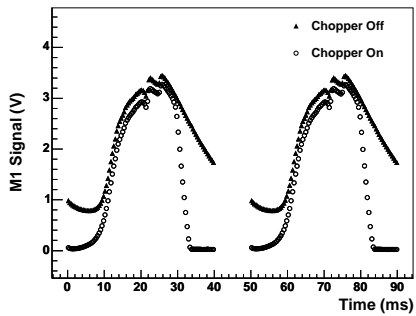}
  \caption{{\em Normalized signal from the first beam monitor downstream
           of the guide exit. The solid triangles show the signal obtained
           from a run where the chopper was parked open. The open circles
           corresponds to a run taken with the chopper running.}} \label{fig:MON1SIG}
\end{figure}

The beam chopper incorporates two blades which rotate independently at
up to $1200$~rpm. The chopper is located $9.38$~m from the surface of
the moderator. Since the flight path is about $21$~m long and the
pulse period is $50$~ms the slowest neutrons that reach the end of the
guide in each pulse have an energy of about $1$~meV. The blades are
coated with a layer of ${\rm Gd_2 O_3}$ which was determined to be
fully absorbing for neutron energies up to $30$~meV. This is used to
block the slow neutrons at the tail end of the time-of-flight spectrum
when either one or both of the blades cover the beam opening. The
diameter of the blades is $1.024$~m and each blade covers
$4.38$~radians of a full circle.  The ability to select only part of
the neutron spectrum is an important tool to control systematic errors
since it provides the ability to effectively polarize, spin-flip and
capture the neutrons.  It prevents the overlapping of very slow
neutrons from a previous pulse with the faster ones from the following
pulse. At a distance of about $22$~m from the moderator, the time
required to fully open or close the beam aperture is $\simeq
4$~ms. During the 2004 commissioning run, the chopper rotation was
phased with the beam pulses such that it began opening at each pulse
onset $(0~{\rm ms})$, was fully open $4$~ms after pulse onset, began
closing about $30$~ms after pulse onset and was completely closed
about $4$~ms later (Fig.~\ref{fig:MON1SIG}).  This allowed us to take
beam-off (pedestal) data for $\simeq 6$~ms at the end of each neutron
pulse, which is needed for pedestal and background studies. The
chopper control feedback loop kept the chopper in phase with the beam
pulse to within $30~{\rm \mu s}$.

\subsection{Beam Monitors}
The experiment uses three parallel plate ion chambers, each with a $12
\times 12~{\rm cm^2}$ active area as beam monitors.  The first
(upstream) monitor is located immediately after the neutron guide
exit. The second monitor is located downstream of the ${\rm ^3He}$
polarizer to allow {\it in situ} absolute beam polarization
measurements to be made. The third monitor is located downstream of
the target and detector array. It was used during the 2004
commissioning run to study the spin flipper efficiency and will be
used to monitor neutron depolarization in the hydrogen target.

\subsection{Neutron Spin Filter}
\begin{figure}[h]
  %\hspace{2cm} %bb=100 0 600 800,clip,angle=270,
  \includegraphics[scale=0.35]{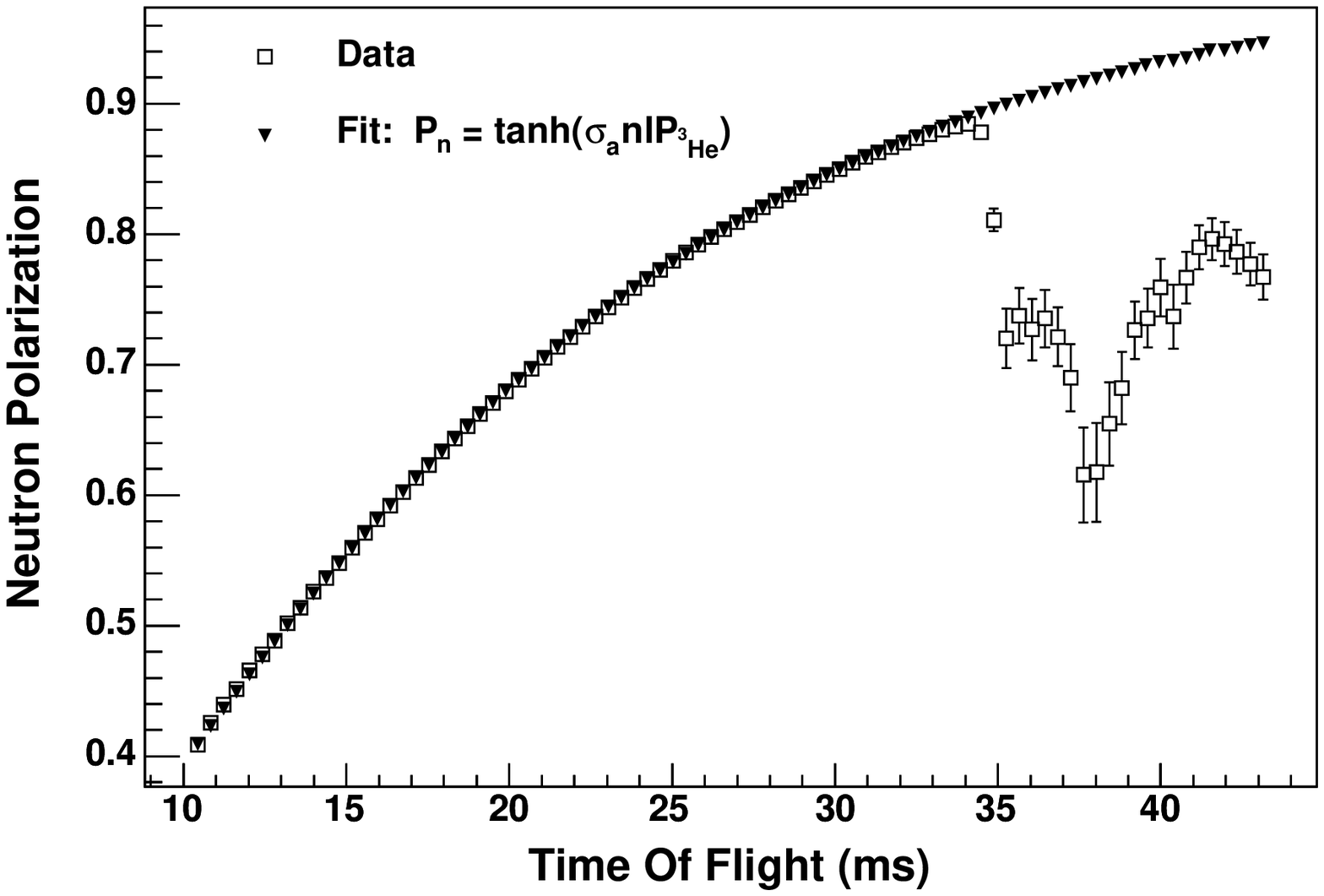}
  \includegraphics[scale=0.35]{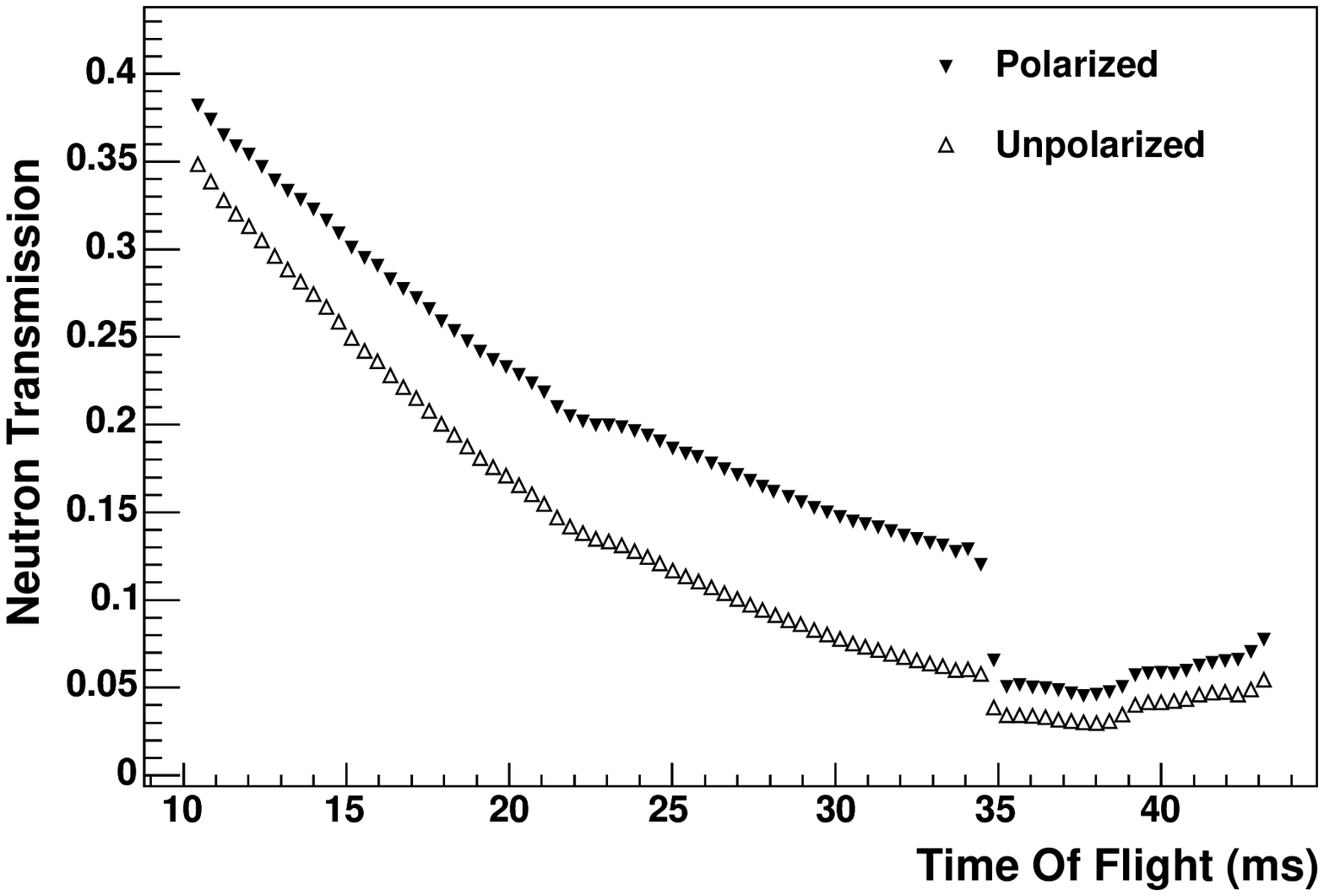}
  \caption{{\em Left: Plot of beam polarization as a function of
           neutron time of flight. Right: Neutron transmission data
           for an unpolarized ${\rm ^3He}$ spin filter cell and a
           polarized cell. The chopper cutoff is completed just below
           35 ms tof.}} \label{fig:NPOL}
\end{figure}

After exiting the neutron guide, the neutrons are spin filtered by
passing through a 12 cm diameter glass cell containing polarized ${\rm
^3He}$. ${\rm ^3He}$ spin filters have a number of desirable
features~\cite{pp:pol}. They have large acceptance angles, do not
require high magnetic fields as is the case with supermirrors and
neutron capture on ${\rm ^3He}$ does not create a $\gamma$-ray
background.  The ${\rm ^3He}$ polarization and therefore the neutron
polarization can be reversed without changing the direction of the
holding field, by adiabatic fast passage of the ${\rm ^3He}$ spin. The
neutron polarization can be measured with $2-3$\% accruracy and
without introducing large magnetic field gradients.  The cross-section
for capture $(\sigma_a)$ of neutrons with spin parallel to the ${\rm
^3He}$ nuclear spin has been measured to be $0.01 \pm 0.03$ of the
total absorption cross-section~\cite{pp:PASS}, but is assumed to be
zero on theoretical grounds. So neutrons with spin anti-parallel to
the ${\rm ^3He}$ nuclear spin are absorbed while those with spin
parallel are mostly transmitted.  If the time of flight (energy) of
the neutrons is known, then the neutron polarization can be determined
directly from the neutron transmission measurements
~\cite{pp:Jones}. For NPDGamma, the figure of merit is the statistical
accuracy that can be reached for a certain running time, which is a
product of both, the neutron transmission and polarization $P_n
\sqrt{T_n}$.  The neutron transmission increases as a function of
energy (decreases as a function of time of flight), whereas the
neutron polarization decreases as a function of energy
(Fig.~(\ref{fig:NPOL})). In the analysis of the data taken during the
2004 commissioning run the neutron polarization was calculated for
each run.  The transmission spectrum obtained for each run was fitted
with $P_{n} = \tanh{(\sigma_a n l P_{He})}$, using a ${\rm ^3He}$
thickness of $nl = 4.84~{\rm bar \cdot cm}$, which was measured with
an unpolarized cell, during the commissioning run.  For each run the
${\rm ^3He}$ polarization was extracted as a fit parameter.

\subsection{Spin Flipper}
In this experiment the asymmetry is measured continuously since the
signals from opposite detectors in a pair are measured simultaneously
for each spin state. However, the efficiency of the $\gamma$-ray
detectors will change slowly due to a number of effects including
temperature and crystal activation and the detector gains cannot be
matched to an accuracy that would allow an asymmetry measurement to be
made for each individual neutron pulse.  In addition, an asymmetry
measurement cannot be made to the required level of accuracy by simply
measuring the signal in a given detector for one spin state and the
corresponding signal in the same detector for the opposing spin state,
made some time later after the neutron spin has been reversed using
the neutron spin filter. This possibility is precluded because of
pulse-to-pulse fluctuations in the beam current. Both situations lead
to false asymmetries.

The primary technique for reducing false asymmetries generated by gain
non-uniformities, slow efficiency changes and beam fluctuations is
fast neutron spin reversal. This allows asymmetry measurements to be
made for opposing detectors, removing sensitivity to beam
fluctuations, and for consecutive pulses with different spin states,
removing the sensitivity to detector gain differences, drifts, and
fluctuations. The asymmetries can then be measured very close together
in time, before significant drift occurs. By carefully choosing the
sequence of spin reversal, the effects of drifts up to second order
can be further reduced. To achieve this fast neutron spin reversal,
the experiment employs a radio frequency neutron spin flipper
(RFSF)~\cite{pp:spfl} which operates according to the principles of
NMR, using a $30$~kHz magnetic field with an amplitude of a few
Gauss. The neutron spin direction is reversed when the RFSF is on and
is unaffected when it is off. During the 2004 commissioning run the
spin flip efficiency was measured to be about 95\% averaged over the
beam cross-section.

\subsection{Liquid Hydrogen Target}
The NPDGamma liquid hydrogen target consists of a cylindrical $20$~l
target vessel containing the liquid hydrogen, surrounded by a vacuum
chamber. The hydrogen itself and the heat radiation shield, located
around the vessel, are cooled by two cryogenic refrigerators.  In the
cooling process, the hydrogen is converted to liquid para-hydrogen,
from its usual state of mostly ortho-hydrogen, to prevent the
depolarization of the neutron spin in the target via spin-flip
scattering. Monte Carlo calculations indicate that the $30$~cm
diameter and $30$~cm long hydrogen vessel is large enough to capture
about 60\% of the incident neutron beam. The beam entrance windows in
the vacuum chamber, radiation shield, and target vessel are as thin as
possible to efficiently transmit the neutron beam and create minimal
prompt capture radiation.

\subsection{Detector Array}
The detector array consists of 48 CsI(Tl) cubes arranged in a
cylindrical pattern in 4 rings of 12 detectors each around the 
liquid hydrogen target.  In addition to the
conditions set on the detector array by the need to preserve
statistical accuracy and suppress systematic effects, the array was
also designed to satisfy criteria of sufficient spatial and angular
resolution, high efficiency, and large solid angle
coverage~\cite{pp:darr}.  To measure $A_{\gamma}$ to an accuracy of $5
\times 10^{-9}$ the experiment must detect at least $ 10^{17}$
\Grs~from \npdg~capture with high efficiency. The average rate
of \Grs~deposited in the detectors for any reasonable run-time is
therefore high. Because of the high rates and for a number of other
reasons discussed in~\cite{pp:darr}, the detector array uses accurate
current mode \Gr~detection. Current mode detection is performed by
converting the scintillation light from CsI(Tl) detectors to current
signals using vacuum photo diodes (VPD), and the photocurrents are
converted to voltages and amplified by low-noise solid-state
electronics~\cite{pp:swb}.

\begin{figure}[h]
  \hspace{4.0cm}
  %bb=a b c d,clip,scale=x    clips the picture to the region from (x=a,y=b) to (x=c,y=d) (x is measured form the left and y from the bottom)
  \includegraphics[scale=0.3]{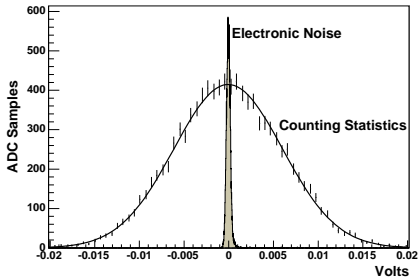}
  \caption{{\em Counting statistics analysis results for a typical
           detector.  The RMS width due to counting statistics is
           compared to the width seen from pedestal runs (electronic
           noise). The fit to the beam on, target in data histogram
           shows an RMS width of $6.1~\pm 0.04$~mV.}}
           \label{fig:CNTSTATS}
\end{figure}

In current mode detection, counting statistics appears as the RMS
width in the sample distribution, due to the fluctuation in the number
of electrons produced at the photo-cathode of the VPD, as a result of
the quasi instantaneous amount of energy deposited in the CsI crystal.
During beam on measurements, the shot noise RMS width is given
by~\cite{bb:DVR}
\begin{equation}
 \sigma_{I_{\mathrm{shot}}} = \sqrt{2qI}~\sqrt{f_B}, \label{eqn:SHN}
\end{equation} 
where $q$ is the amount of charge created by the photo cathode per
detected \Gr, $I$ is the average photo-current per detector and
$f_B$ is the sampling bandwidth, set by the $0.4$~ms time bin width in
the time of flight spectrum.  Figure~\ref{fig:CNTSTATS} shows the RMS
width for a typical detector, as seen at the preamplifier output. The
width from counting statistics is compared to the RMS width seen for
beam-off electronic noise.

To ensure a timely and accurate measurement of the \Gr~ asymmetry the
detector array must operate at counting statistics which requires the
RMS width from electronic noise to be significantly smaller than the
width observed from real events. For a given $q$ and $I$, the RMS
width expected from the detector design criteria at the preamplifier
output is $5.7~\pm 0.3$~mV. The error on the expected width is
dominated by the accuracy to which we know the efficiency (number of
photo-electrons per MeV) of the detector.  Using the known proton
current and an appropriate Monte Carlo model for neutron capture on
\BC~ the number of photons entering a detector, per time bin, is
$3\times 10^{4}$. The corresponding RMS width expected due to neutron
counting statistics is $\simeq 6 \pm 0.5$~mV.

\section{2004 Commissioning Targets Asymmetries} \label{scn:ASYM}

To determine the parity violating asymmetry in neutron-proton capture
to the proposed accuracy, any possible {\em false} asymmetry from
neutron capture on other materials must be measured. These asymmetries
form a background which introduces a shift in the measured \npdg~
asymmetry if they are non-zero and, at the very least, produce a
dilution of the asymmetry, even if they are zero. The degree of shift
or dilution in the asymmetry is proportional to the size of the
background signal, relative to the signal of interest. We refer to
these {\em false} asymmetries as {\it neutron capture related or
induced} systematic effects. There are also {\it instrumental}
systematic effects which arise due to changing equipment properties
which may be correlated with the neutron spin. Some of the possible
instrumental systematic effects have been briefly mentioned earlier
and those related to the detector array in particular are discussed in
detail in~\cite{pp:darr}. A more detailed discussion of systematic
effects related to neutron capture and scattering is provided in
~\cite{th:ger}. It is difficult to model or calculate the level of
parity-violation in these targets and to establish an upper level of
their contribution it must be measured.

\subsection{Asymmetry Definition}

Due to the $10$~G holding field, surrounding the experimental
apparatus, the neutrons are polarized vertically after leaving the
$^3$He spin filter. While taking hydrogen data, the parity violating
asymmetry in n-p capture is therefore seen in a difference of the
number of $\gamma$-rays going up and down. For the ideal experiment,
the $\gamma$-ray cross section is proportional to $Y = 1 +
A_{\gamma}~\cos{\theta}$, where $\theta$ is the angle between the
neutron polarization and the momentum of the emitted photon. A third
term is introduced if a left-right (LR) asymmetry exists $Y =
1+A_{\gamma}~\cos{\theta}+A_{\gamma,LR}~\sin{\theta}$. However, as
discussed earlier, the basic expression for the \Gr~ yield is modified
due to limitations in the properties of the experimental apparatus and
interaction of neutrons with elements other than hydrogen.  In
calculating the final combined asymmetry one asymmetry was calculated
for each detector pair and time bin and over any valid sequence of 8
macro pulses with the correct neutron spin state pattern.  A so-called
valid 8 step sequence of spin states is defined as
($\uparrow\downarrow\downarrow\uparrow\downarrow\uparrow\uparrow\downarrow$).
This pattern suppresses first and second order gain drifts within the
sequence.  The measured ({\em raw}) asymmetry $(A^{j,p}_{raw})$ for
each detector pair $(p)$ can be extracted by forming a ratio of
differences between detectors in a pair divided by their sum. After
all correction factors have been applied, the final physics asymmetry
for a given detector pair, spin sequence $(j)$, and neutron time of
flight $(t_i)$ is given by
\begin{eqnarray}
  \left(A^{j,p}_{UD}(t_i) + \beta A^{j,p}_{UD,b}(t_i)\right)\langle G_{UD}(t_i) \rangle &+& 
  \left(A^{j,p}_{LR}(t_i) + \beta A^{j,p}_{LR,b}(t_i)\right)\langle G_{LR}(t_i) \rangle \nonumber \\ 
  &=& \frac{\left(A^{j,p}_{raw} - A^p_g A_f(t_i) - A^p_{noise}\right)}{\epsilon(t_i)P_n(t_i)S(t_i)} \nonumber \\
  \label{eqn:TBPHASY}
\end{eqnarray}
Where the background asymmetries $(A^{j,p}_{UD,b}, A^{j,p}_{LR,b})$
and the relative signal level $(\beta)$ from the elements that cause
them must be measured in auxiliary measurements. The relative, target
out, background levels for the various targets were $7\% \leq \beta
\leq 17\%$, depending on the shielding collimation and target
geometries. Asymmetries from target out runs were measured to be zero.
$A^p_g$ is the gain asymmetry between the detector pair and $A_f(t_i)$
is the asymmetry from pulse to pulse beam fluctuations. The neutron
energy and detection efficiency weighted spatial average detector
cosine (up-down asymmetry) with respect to the vertical is given by
$\langle G_{UD}(t_i) \rangle \simeq cos(\theta)$, while the detector
sine (left-right asymmetry) is given by $\langle G_{LR}(t_i) \rangle
\simeq sin(\theta)$. Also included are the correction factors due to
the neutron beam polarization $(P_n(t_i))$, the spin flip efficiency
$(\epsilon(t_i))$ and the level of beam depolarization in the target
$(S_n)$. The beam depolarization for the targets we report on here was
modeled and values for $(S_n)$ range from $0.95$ to $1$. It is
important to realize that signal fluctuations that are not correlated
with the switching of the neutron polarization direction will average
out and don't contribute to any asymmetry. It is, however, essential
that these signals have an RMS width that is small compared to the RMS
width in the asymmetries of interest (driven by counting statistics)
so that they do not dilute the result and are averaged to zero quickly
compared to the time it takes to measure the asymmetry to the desired
accuracy. The product of the gain and beam fluctuation asymmetries was
measured to be consistent with zero with a statistical error that was
typically two orders of magnitude smaller than the error on the raw
asymmetry (see RMS width in table ~\ref{tbl:FNASYVALS}).  Possible
false asymmetries due to electronic pickup and possible magnetic field
induced gain changes in the detector VPDs have previously been
measured and are consistent with zero to within $5 \times
10^{-9}$~\cite{pp:darr}.

The detector pair physics asymmetries as represented by
eqn.~\ref{eqn:TBPHASY} can then be combined in error weighted averages
over the neutron time of flight spectrum to form a single asymmetry
for the entire detector array for a single sequence of beam pulses.
If beam intensity levels are monitored to be reasonably stable over the
measurement time these sequence asymmetries can be histogrammed.
Typical run lengths were $\sim 8.3$ minutes and included 10000 beam
pulses or 1250 8-step sequences and the asymmetry measurements performed
so far usually extended over several hundred runs.
\begin{figure}[ht]
  \hspace{0cm}
  \includegraphics[scale=0.32]{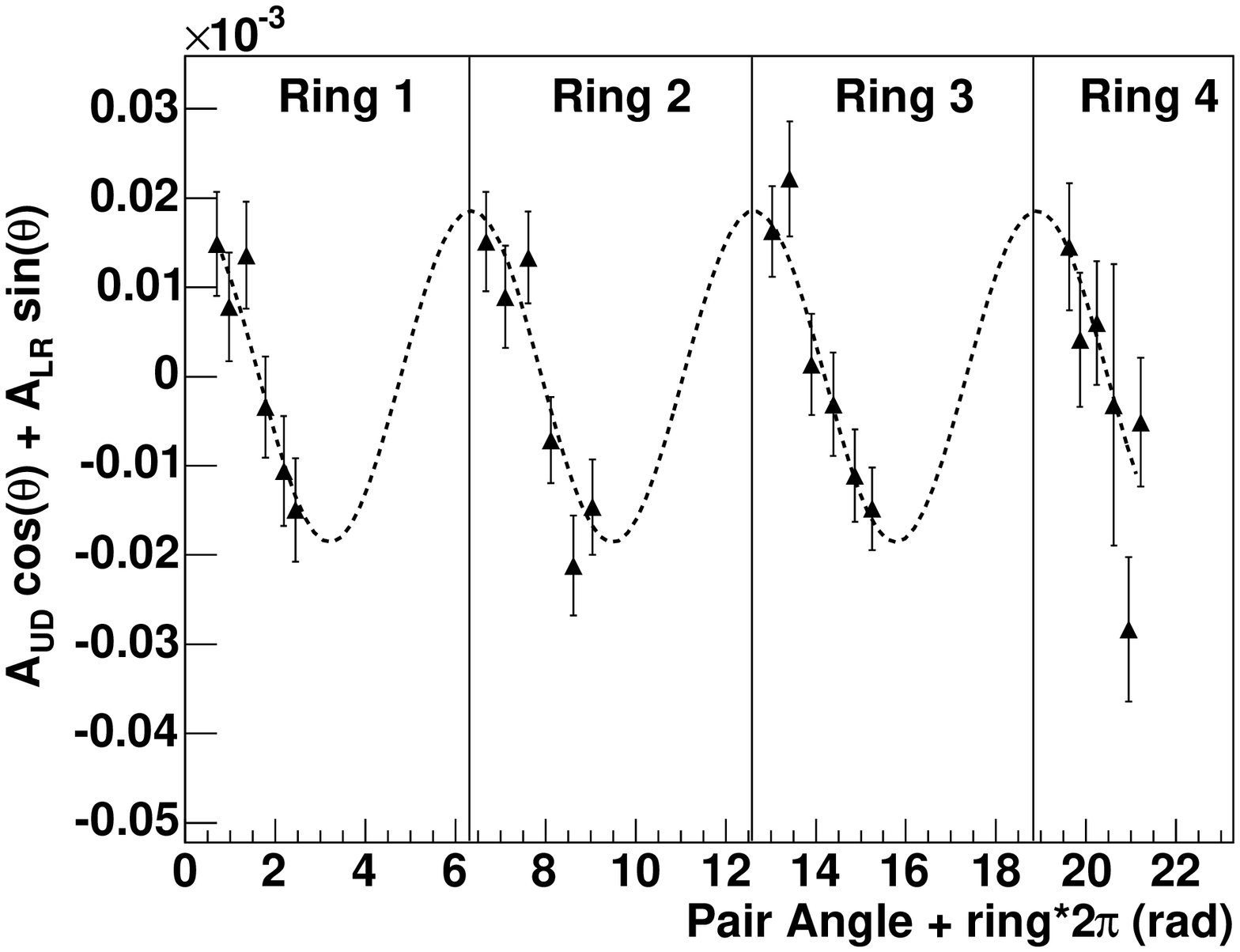}
  \includegraphics[scale=0.35]{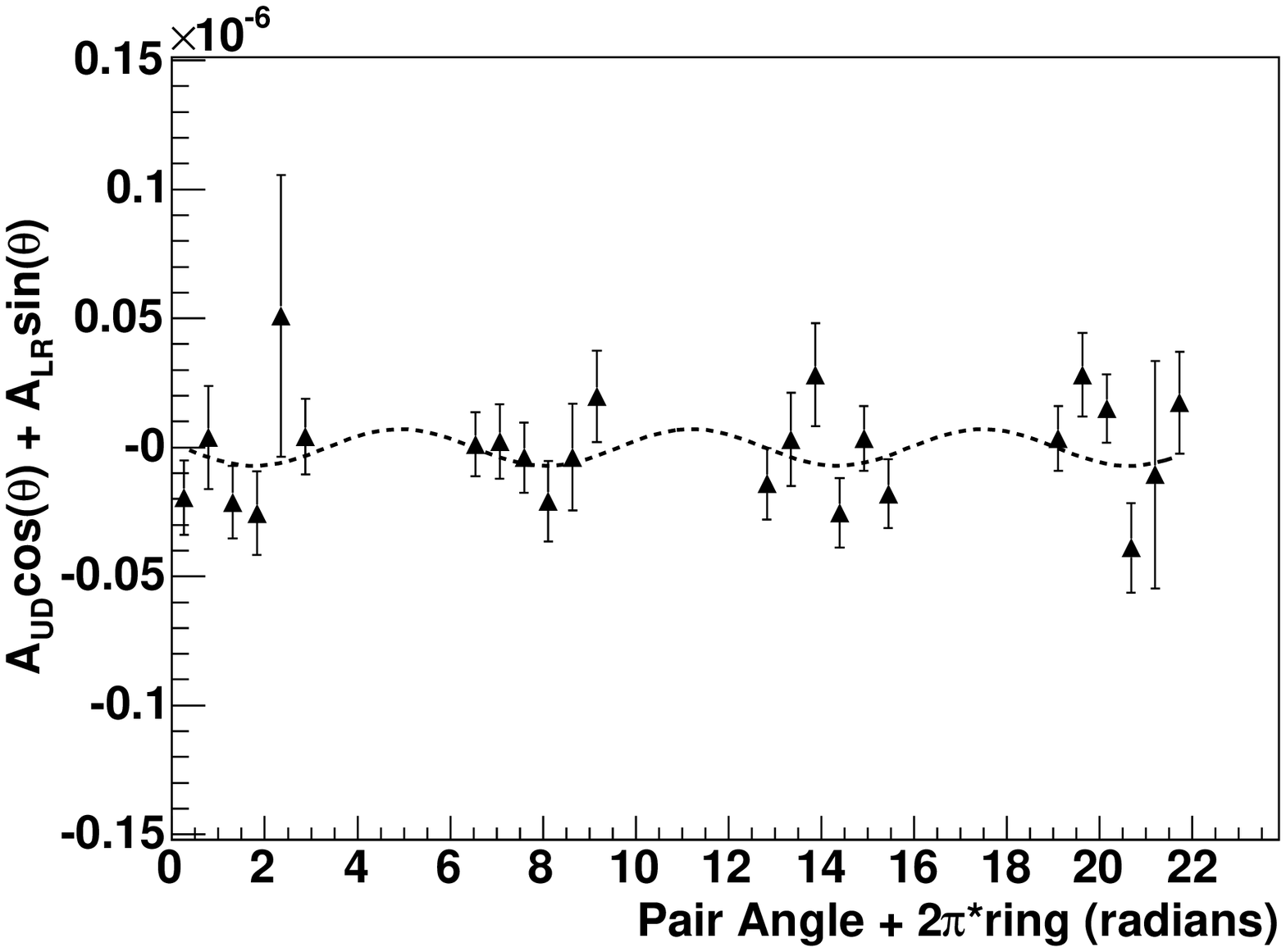}
  \caption{{\em Left: \CCL~asymmetries for each pair, plotted versus angle of the
           first detector in the pair w.r.t the vertical. The total
           array asymmetry is extracted from the
           fit. Right: Noise asymmetries.}}\label{fig:CCLPRASYS}
\end{figure}

\subsection{Results}

The known asymmetry in \CCL~was used to verify that a nonzero
asymmetry can, in fact, be measured with this experimental setup. The
\CCL~asymmetry was also used to verify the geometrical dependence of
the pair asymmetries. For this purpose, each of the 24 pair asymmetries,
extracted from the histogrammed 8-step sequence asymmetries from all
data obtained with that target, were multiplied by its mean geometry
factor and plotted versus its corresponding mean error.  The resulting
graph is shown in Fig.~\ref{fig:CCLPRASYS}. The fit function used to
extract the total array asymmetry is $A_{UD}\cos{\theta} +
A_{LR}\sin{\theta}$. The ${\rm ^{35}Cl}$ asymmetries obtained from the
\CCL measurements were previously measured by this collaboration and
others.  M. Avenier and collaborators~\cite{pp:AVENIER} found an
Up-Down asymmetry of $(-21.2 \pm 1.7)\times10^{-6}$.  The results of
the asymmetry measurements performed during 2004 commissioning run are
summarized in table~\ref{tbl:FNASYVALS}. Systematic errors from the
correction factors discussed above are less than 10\% and are scaled by 
the asymmetry.
\begin{table}[hb]
\begin{center}
  \begin{tabular}{l|r|r|r} 
    \multicolumn{4}{c}{${\bf Asymmetries~and~RMS~width}$} \\ \hline
                 &{\bf Up-Down}                            &{\bf Left-Right}                         &{\bf RMS width (typ.)}\\ \hline
    Al           & $\left(-0.02 \pm  3\right)\times10^{-7}$& $\left(-2   \pm  3\right)\times10^{-7}$ & $1.2\times10^{-3}$   \\   
    \CCL         & $\left(-19   \pm  2\right)\times10^{-6}$& $\left(-1   \pm  2\right)\times10^{-6}$ & $1.0\times10^{-3}$   \\  
    \BC          & $\left(-1   \pm  2\right)\times10^{-6}$ & $\left(-5   \pm  3\right)\times10^{-6}$ & $0.7\times10^{-3}$   \\ 
    Cu           & $\left(-1   \pm  3\right)\times10^{-6}$ & $\left( 0.3 \pm  3\right)\times10^{-6}$ & $1.0\times10^{-3}$   \\  
    In           & $\left(-3   \pm  2\right)\times10^{-6}$ & $\left( 3   \pm  3\right)\times10^{-6}$ & $0.4\times10^{-3}$   \\
    Noise (add.) & $\left( 2   \pm  5\right)\times10^{-9}$ & $\left(-7   \pm  5\right)\times10^{-9}$ & $2.0\times10^{-6}$   \\
    Noise (mult.)& $\left( 3   \pm  7\right)\times10^{-9}$ & $\left(-9   \pm  7\right)\times10^{-9}$ & $0.2\times10^{-3}$   \\
    Beam $\times$ Gain & 0                                 & 0                                       & $1.0\times10^{-5}$   \\
  \end{tabular} 
\end{center}
  \caption{{\em Up-Down and Left-Right asymmetries for the target
           materials used during the 2004 commissioning run. Stated
           errors are statistical only. Systematic errors are less
           than 10\% and are scaled by the asymmetry. The RMS widths
           are taken from histograms with single 8-step sequence
           asymmetries for a detector pair as individual
           entries. There are no up-down or left-right beam and gain
           asymmetries because they are independent of the direction
           of neutron polarization and \Gr~
           emission.}}\label{tbl:FNASYVALS}
\end{table}

\section{Conclusion}

The NPDGamma experiment successfully completed a commissioning run in
April 2004. It was shown here that each component in the experiment,
except the hydrogen target, was commissioned during the 2004 run cycle
and each component performed as designed.  To establish the level of
false asymmetries that may be present for the hydrogen production
runs, several measurements were performed.  Possible false asymmetries
due to instrumental systematic effects involving the detector array
and spin flipper were measured to be zero at the $5\times10^{-9}$
level. Asymmetries due to beam fluctuations were measured using the
beam monitors. The beam asymmetry enters into the main data asymmetry
as a product with the detector pair gain asymmetry with a combined RMS
width of $10^{-5}$, which is negligible. False asymmetries due to
neutron capture on materials other than hydrogen were measured for
Al, Cu, In and \BC. These asymmetries were found to be consistent
with zero. It is clear from the results obtained so far that NPDGamma
incorporates a powerful experimental setup that can be used to measure
very small parity violating asymmetries and that the experiment meets
all criteria needed to perform a successful measurement of the weak
parity-violating $\gamma$ asymmetry with an accuracy of $5\times
10^{-9}$ in the neutron capture reaction \npdg\ .

%\bibliography{dettest}

\end{document}